\documentclass[11pt,a4paper]{article}

\usepackage[margin=1in]{geometry}
\usepackage{graphicx}
\usepackage[caption=false,font=footnotesize]{subfig}
\usepackage{amsfonts,amsmath,amsthm,amssymb,float,url,bbm}
\usepackage[utf8]{inputenc}
\usepackage[T1]{fontenc}
\usepackage{xcolor}
\usepackage{booktabs}
\usepackage{authblk}
\usepackage{orcidlink}
\usepackage{microtype}
\usepackage{booktabs}
\usepackage{multirow}

\hypersetup{
    pdfstartview=XYZ,
    bookmarks=true,
    colorlinks=true,
    linkcolor=blue,
    urlcolor=blue,
    citecolor=blue,
    linktocpage=true,
    hyperindex=true
}

\graphicspath{{./figures/}{./pnr/}}

\renewcommand{\phi}{\varphi}

\title{Minimum RISC-V:\\ Maximum FPGA Efficiency}

\author[1]{Martin Langhammer\,\orcidlink{0000-0001-8206-2077}}
\author[2]{Bogdan Pasca\,\orcidlink{0000-0002-5454-4375}}
\author[3]{Gregg Baeckler\,\orcidlink{0009-0002-1212-8071}}

\affil[1]{Altera Corporation, London, UK\\
\texttt{martin.langhammer@altera.com}}
\affil[2]{Altera Corporation, Paris, France\\
\texttt{bogdan.pasca@altera.com}}
\affil[3]{Altera Corporation, San Jose, USA\\
\texttt{gregg.baeckler@altera.com}}

\newcommand{\na}{\multicolumn{1}{c}{--}}

\date{}

\begin{document}
\maketitle

\begin{abstract}
RISC-V processors are now widely available for FPGAs, with most FPGA vendors replacing their proprietary soft-CPU architectures with this instruction-set architecture (ISA).
Although contemporary FPGAs are increasingly resource- and feature-rich, a soft processor system---including memory---often occupies a disproportionately large footprint and delivers only modest performance relative to the potential of the target device.
In this work we first design a lightweight but deeply pipelined RISC-V core that can run above \textbf{700\,MHz} in a mid-range device and that is capable of retiring one instruction per clock under ideal conditions.
We then develop and demonstrate a new optimization method that compiles the final application binary and a specialized processor instance \emph{contemporaneously}.
The method expands each instruction into an application-specific encoding, removes unused datapath and decode logic during synthesis, and yields cores that remain assembler-compatible with the original program while no longer being binary-compatible with a general-purpose RISC-V.
We implement example designs on RV32I (base integer), RV32M (integer multiply), and RV32F (single-precision floating-point) configurations.
Our optimized RISC-V is as small as \textbf{170 ALMs} and closes timing at \textbf{775\,MHz} in an Altera Agilex~5 device.
Finally, we show that these processors can be packed into a deterministic area and stepped-and-repeated continuously without material impact on performance, enabling predictable multi-core density planning.
\end{abstract}



\section{Introduction}

The RISC-V ISA~\cite{riscv-isa-2019} is increasingly common in both research and commercial contexts, and is now directly supported by most FPGA vendors~\cite{microchip_riscv,efinix_sapphire,Lattice_RISCV} in soft and, increasingly, hard processor offerings.
The Altera Nios~II~\cite{Nios} processor has been replaced by Nios~V~\cite{NiosV}; Xilinx/AMD has similarly replaced MicroBlaze~\cite{Microblaze} with MicroBlaze~V~\cite{MicroBlazeV}.
Continued investment by FPGA vendors in soft CPUs underscores their utility, even on devices that already contain hard processor subsystems~\cite{AgilexARM,Zync}, typically ARM-based.

There are many good reasons to adopt RISC-V---notably standard tooling and a growing software ecosystem.
However, RISC-V was not architected specifically for FPGAs, and several aspects of the ISA map poorly onto FPGA fabrics.
Some of these mapping issues become concrete in Section~\ref{sec:lightweight}, where we detail the design of a lightweight RV32I core.

Many FPGA applications demand a small, inexpensive device in which a full-featured CPU would be too expensive.
Often these cases involve only a small code size, with applications such as motor control or complex state machines.
In other cases the CPU may only monitor or manage dedicated IP blocks, or perform a slower offline task that is too complex (or large) to implement conveniently in RTL.
The processor then becomes an expensive, unwieldy part of the solution---out of proportion to its functional contribution.

In this work we introduce an \emph{application-specific} soft-processor methodology.
The design can be developed and iterated with standard RISC-V tools.
In the final stage, an additional processing step compiles a processor version that supports only the associated software application.
This can reduce processor size by 50\% to 70\%, partially by removing portions of the datapath that are unused by the remaining instruction mix.
Although we start from the RV32I instruction set, the specialized core will not run a generic RISC-V program: forwarding has been removed (and replaced with NOP insertion) to save logic.

The resulting optimized processor is not only smaller in resource count, but also uses the FPGA more efficiently.
Small application spaces---for example an RV32I core with roughly 500 assembly instructions and around 1000 32-bit data words---can be compiled into a structure with a balanced use of resources, i.e., where the ratio of memory to logic mirrors that of the FPGA.
This enables very efficient placement of the CPU onto the device.

\paragraph{Contributions.}
We make the following contributions:
\begin{itemize}
    \item We develop and evaluate an optimization method for the RISC-V ISA that greatly reduces logic and other resources for application-specific code examples.
    The method is not RISC-V-specific and can be applied to other processor architectures.
    \item We benchmark the effect of processor complexity features---notably forwarding and hardware interlocks---against performance and code size.
    \item We describe the true cost of optimized processors as \emph{footprint} on the FPGA, rather than as an unstructured list of resources.
    This cost measure is also applicable to other FPGA designs.
\end{itemize}

\paragraph{Paper organization.}
Section~\ref{sec:related} surveys contemporary FPGA soft RISC-V cores and related optimization approaches.
Section~\ref{sec:lightweight} presents our lightweight baseline RV32I microarchitecture.
Section~\ref{sec:optimize} describes the contemporaneous application/processor optimization flow.
Sections~\ref{sec:quicksort}--\ref{sec:rv32f} evaluate the method on RV32I Quicksort, an RV32M FFT profile, and an RV32F Gaussian matrix solver.
Section~\ref{sec:results} analyzes placement footprint and multi-core packing.
Section~\ref{sec:conclusions} concludes.

\begin{figure*}[t]
    \centering
    \includegraphics[width=0.99\textwidth]{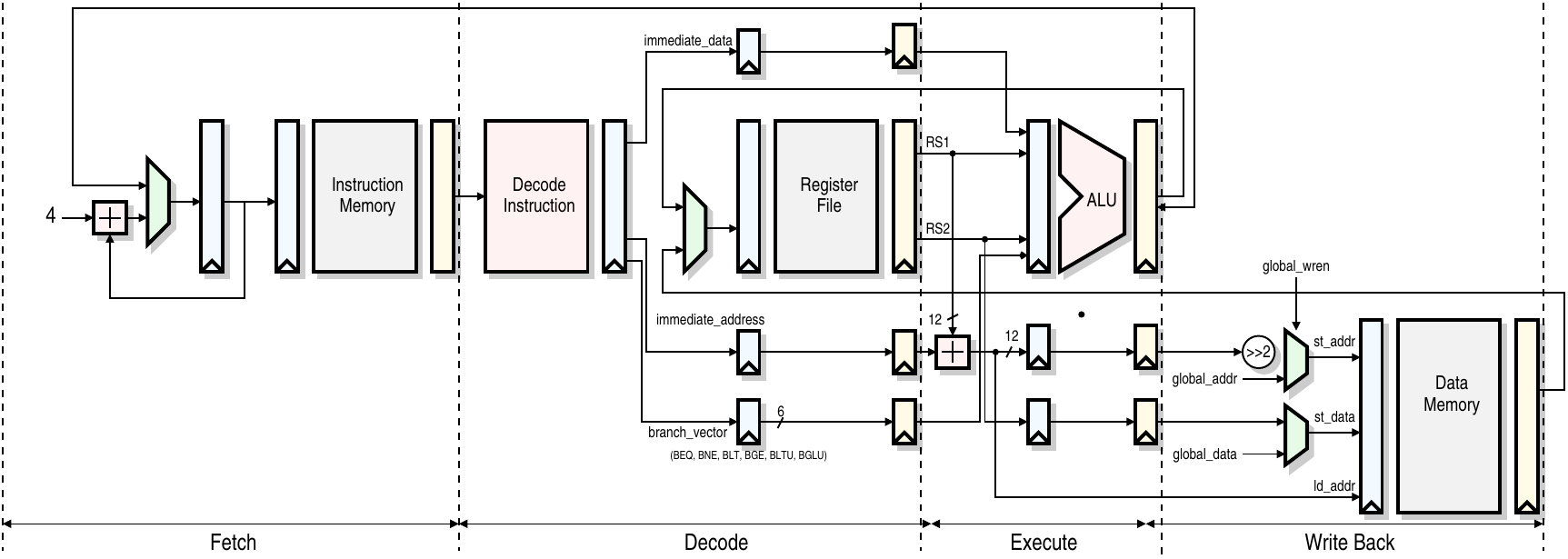}
   \caption{Lightweight RISC-V cores. Yellow register stages are skipped in the 5-stage pipeline architecture.}
   \label{fig:riscv_main_xfig}
\end{figure*}

\section{Current FPGA RISC-V Cores}
\label{sec:related}

In this section we examine the current state of soft RISC-V processors from several perspectives: standard vendor IP offerings, processor optimizations, and multi-processor implementations.
All of these themes reappear in our specialized-processor results.

FPGA vendors typically offer a RISC-V processor family with many feature and performance options.
Most of these generators produce cores in the 1000--2000 LUT range that close timing at only a fraction of the device potential---commonly around 300\,MHz for a five-stage RISC-V.
The Lattice~\cite{Lattice_RISCV} core is specified at only 100\,MHz, but that figure is for a low-power family with a 350\,MHz clock tree~\cite{latticesemi2020certusnx}.
Independent analysis of the Xilinx/AMD MicroBlaze~V~\cite{Adiuvo} shows logic resources ranging from 823 to 1765 6-LUTs for a minimal RV32I configuration, depending on pipeline depth (3 to 8 stages) and selected performance features.
A comparison of a large set of contemporary soft RISC-V research cores reported even larger resource requirements, often multiple thousands of LUTs~\cite{Tapasco}.
A CPU core requiring ``only'' 1000 LUTs can still be a very expensive component in a low-cost device, especially once footprint impact---placement density including memory---is considered.

A number of optimization approaches have been reported.
Phalanx GRVI~\cite{Phalanx,Hoplite} is a large array of RISC-V RV32I cores in which expensive ALU components (such as 32-bit barrel shifters) are shared among multiple processors.
In other work, an array of multi-threaded RISC-V cores~\cite{Dirk_FPL2025} configures each CPU as a barrel processor (multiple independent threads) with a varying number of threads and an equivalent pipeline depth, achieving clock frequencies up to 737\,MHz.
Unfortunately, neither project reported application-program results that would allow end-to-end cost/performance comparison.

System performance for other soft RISC-V arrays~\cite{Diana,Agiler,HERO} has been reported, but typically with low clock rates or poor FPGA density.
For example, a 36-processor system comprising nine clusters of four processing elements (PEs), each PE a RISC-V RV32IMC RISCY~\cite{RISCY} core, consumed 395K 6-LUTs out of 1186K available.
That corresponds to only 33.2\% logic efficiency, and the published floorplan contains significant areas of unused---and generally unreachable---logic.
Such results illustrate a recurring issue in FPGA system design: resource counts alone do not capture whether a design can be packed densely or clocked near the fabric limit.

\section{A Lightweight FPGA RISC-V Core}
\label{sec:lightweight}

\begin{figure}[t]
    \centering
    \includegraphics[width=\columnwidth]{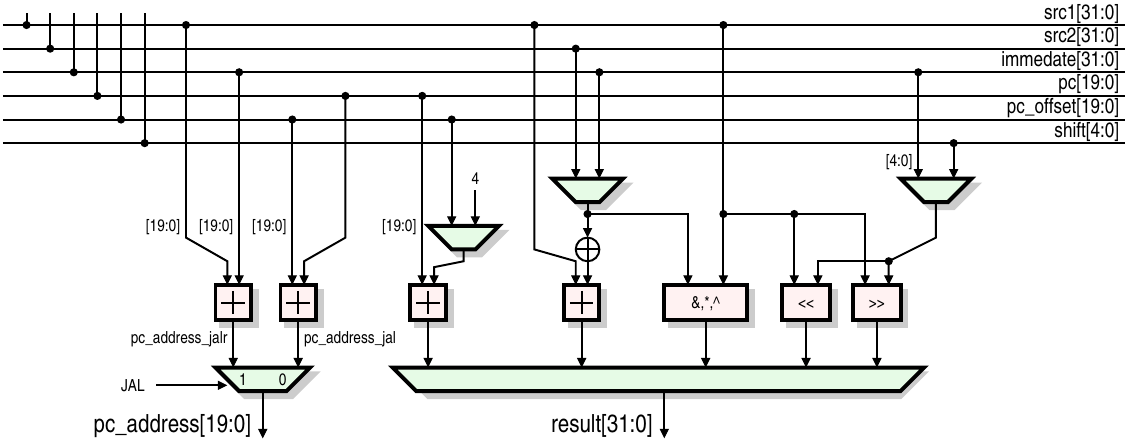}
   \caption{Detail of the ALU section of our lightweight RISC-V processor.}
   \label{fig:riscv_alu}
\end{figure}

Most FPGA RISC-V cores require substantially more resources than well-established vendor-specific soft CPUs such as Nios~II~\cite{Nios} (Altera) and MicroBlaze~\cite{Microblaze} (Xilinx).
To explore optimizing RISC-V for FPGA, we first develop a lightweight RISC-V RV32I core.
Our optimization research initially targets single-threaded operation, so we do not support multi-threaded (\texttt{FENCE*}), Control and Status Register (\texttt{CSR*}), or operating-system support instructions such as \texttt{ECALL}.
All other instructions in the RV32I instruction set can be executed.
All reads and writes have word (32-bit) precision; byte and half-word accesses therefore require multiple instructions and must be enabled as a compiler option.

Our goal was to build the smallest practical processor while approaching the maximum clock rate of the FPGA.
This required a deep pipeline, which in turn increased the potential for data hazards.
For reasons of both simplicity and cost we did not include forwarding or hardware interlocks; hazard avoidance is performed with software NOPs.
While NOPs can significantly increase program size, we show that this is not a large cost for the application-specific setting considered here.

The processor executes all instructions in a single issue cycle, so the clocks-per-instruction (CPI) can be as low as 1.0 in the ideal case, although this rarely occurs in practice.
In addition to NOPs that handle data hazards, branch delay slots are long for a simple processor without branch prediction (our processor automatically flushes the instruction pipe on a taken branch).
Our goal is that absolute wall-clock performance should remain competitive with other FPGA RISC-V cores because of the high clock rate of our design, and that the \emph{normalized} cost---performance adjusted for FPGA footprint---should be better.

We target mid-range and low-cost devices, specifically the Altera Agilex~5~\cite{alteraagilex5datasheet} and Agilex~3~\cite{altera2026agilex3pb} families.
The processor also maps directly to other Altera devices with a similar logic and memory structure.
The core is written in generic RTL and should therefore be synthesizable to FPGAs from other vendors, although we did not evaluate that path in this work.

\subsection{Pipeline organization}

Our base core uses an 8-stage pipeline organized somewhat differently from the standard textbook five-stage architecture.
A block diagram is shown in Figure~\ref{fig:riscv_main_xfig}.
The Instruction Memory (I-MEM) is implemented in Altera M20K~\cite{Agilex_Memory} memories, which support embedded input and output pipeline stages and therefore require no registers in the soft logic fabric.
The decode stage is shallow---in most places only one level of logic deep---but adding a pipeline stage here is essentially free, using registers already present in the ALM~\cite{Agilex_Logic} (the configurable 6-LUT that comprises the Agilex soft logic).
The register file is implemented with two parallel M20K memories configured in simple dual-port mode (1R--1W), which can operate at the device maximum clock frequency.
Again we take advantage of the embedded input and output registers of the memory block.

The Data Memory (D-MEM) is also mapped to M20K memories.
A single M20K can be configured as a $512\times 32$ block; two blocks are sufficient for all application examples used in this work.
Data address offsets are applied by dedicated adders immediately preceding the D-MEM---as opposed to the typical practice of using the ALU---because a physical adder exists at every level of soft logic in the FPGA.
All memories are physically arranged in columns.
A lightweight processor, which typically contains only a small number of M20Ks, must route from the output of one memory (the register file) to the input of an adjacent memory (or memories).
Paths from the D-MEM and ALU outputs are multiplexed into the register-file write port.
There is a single unregistered logic level here, again crossing from one side of a memory column to the other.
The address-offset adder creates an almost identical routing path.
These paths, together with the ALU, are potential critical paths.
Placement structure is revisited in Section~\ref{sec:results}.

\subsection{ALU and delay structures}

The largest block in the processor is the ALU, shown in Figure~\ref{fig:riscv_alu}.
Each function is instantiated as a separate entity at the same logic level.
While this creates a wide structure, it is also very shallow, yielding a short path from the ALU input to its output register.
The combinatorial delays of the three critical paths in the processor are approximately equal: (i)~internal ALU; (ii)~register file to D-MEM through the address-offset adder; and (iii)~D-MEM/ALU back to the register file.
Although some paths inside the ALU contain several levels of logic (e.g., shifters), this is mitigated by grouping the ALU in one local logic region rather than routing across a memory column.
For some highly optimized cases the ALU consumes over 60\% of the remaining logic and is therefore distributed across the core; even then the floorplan is so compact that relative placement is less critical.

Because of both the ISA and our particular implementation, the architecture requires a large number of deep delay chains to align decoded instructions with their consumers.
Many of these delays can be absorbed by the secondary registers in the ALM.
Each ALM contains four registers: two typically connected directly to ALM logic for pipelining (primary), and two others that can reduce fan-out or---if local interconnect is available---implement additional delays (secondary).
In many Altera devices, Hyper-Registers~\cite{ChromczakAgilex} are also embedded in the routing fabric and are automatically used by Quartus for pipelining.
Hyper-Registers do not support resets and are therefore used only when processor delay structures are implemented without them.

\subsection{Five-stage variant}

We also created a 5-stage pipeline version of the CPU based on the 8-stage design.
Figure~\ref{fig:riscv_main_xfig} shows in yellow which pipeline levels were removed.
Functionality is unchanged; the three removed stages are the I-MEM output register, the register-file output register, and the D-MEM output register (the ALU output register at the same level as the D-MEM is also removed).
This change removes two cycles from the worst-case data-hazard condition of the 8-stage processor (a maximum of two NOPs instead of four after any instruction) and three cycles from the branch delay slot.
Clock rate decreases by about 35\%, but under some conditions system performance may be comparable: if an application contains many branches or many data hazards, wall-clock time of the two processors can be similar.
The 8-stage processor runs much faster than typical FPGA designs (e.g., the average speed of designs presented at a recent major FPGA conference~\cite{fpl2024proceedings} was 275\,MHz); both of our processors close timing well above the rates of systems into which they may be incorporated.

Table~\ref{tab:full_resources} reports the resource mix for the 8-stage and 5-stage processors.
Removing resets from delay structures reduces logic use by approximately 5\%, as ALM registers can be replaced by Hyper-Registers or MLABs~\cite{Agilex_Memory} (groups of 10 ALMs configured as $32\times 20$ FIFOs).


\begin{table*}[t]
  \centering
  \caption{Lightweight RISC-V cost when compiled to Agilex A5EC013AB23 devices.}
  \label{tab:full_resources}
  \footnotesize
  \setlength{\tabcolsep}{3pt}
  \renewcommand{\arraystretch}{1.1}
  \begin{tabular}{@{}l l @{\hspace{6pt}} c c @{\hspace{6pt}} c c c c @{\hspace{6pt}} c c c@{}}
    \toprule
    & & \multicolumn{2}{c}{\textbf{ALMs}} & \multicolumn{4}{c}{\textbf{Registers}} &
    \multicolumn{3}{c}{\textbf{FMax (MHz)}} \\
    \cmidrule(lr{4pt}){3-4} \cmidrule(lr{4pt}){5-8} \cmidrule(l{4pt}){9-11}
    \textbf{Processor} & \textbf{Synthesis} &
    \textbf{Total} & \textbf{ALU} &
    \textbf{Total} & \textbf{Primary} & \textbf{Secondary} & \textbf{Hyper} &
    \multicolumn{3}{c}{\textbf{Speed Grade}} \\
    & \textbf{Option} & & & & & & & \textbf{-1} & \textbf{-2} & \textbf{-3} \\
    \midrule
    \multirow{2}{*}{RISC-V 8-Stage}
      & Default            & 447 & 250 & 914 & 751 & 152 &  11 & 658 & 639 & 511 \\
      & No Reset on Delays & 424 & 248 & 963 & 690 & 126 & 147 & 706 & 640 & 525 \\
    \midrule
    RISC-V 5-Stage
      & Default            & 493 & 248 & 762 & 634 & 128 &   0 & 428 & 383 & 301 \\
    \bottomrule
  \end{tabular}
\end{table*}

\section{Optimization Approach}
\label{sec:optimize}

\begin{figure*}[t]
    \centering
    \includegraphics[width=0.99\textwidth]{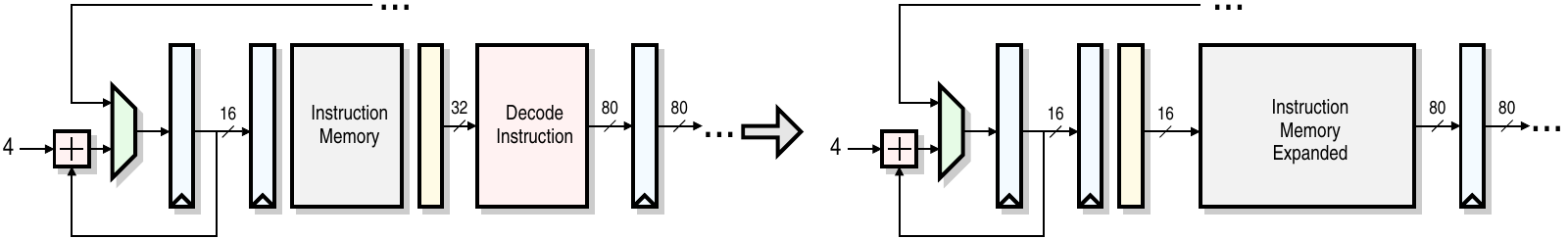}
   \caption{Optimization step: architecture after expanding and collapsing the instruction word.}
   \label{fig:riscv_optimize}
\end{figure*}

The lightweight RISC-V processor is already much smaller and faster than many previously reported cores, albeit at the expense of CPI due to the lack of forwarding and the increase in program size from software hazard mitigation.
In an FPGA context, however, no two processor implementations need be identical, and there can be significant advantage in further specializing the processor---i.e., creating application-specific soft CPUs.
Some target applications require a single complex algorithm (such as motor control) or replace complex state machines for which RTL programming is cumbersome.

We profiled a number of applications; sorting, FFT, and matrix solving are presented as detailed examples later in this paper.
For many applications only a subset of the instruction set is used.
That observation simplifies instruction decode and, more importantly, can be used to simplify the processor datapath itself.

\subsection{Flow}

We start with the application, or a portion of it, written in C.
We compile to a bare-metal RISC-V processor using gcc~15.2.0 on Compiler Explorer~\cite{Godbolt}.
Next, using Claude Code~\cite{claude_ai}, we insert NOPs into the binary and adjust all branch addresses as required.

The standard RISC-V binary is 32 bits wide; different fields map to different ranges depending on the instruction type.
We extract each field type---register sources and targets, and the various immediate values (PC offsets, address offsets, immediate operands, and shift amounts)---individually.
Each instruction, together with an instruction-type identifier where used, is enumerated separately in a one-hot field.

For any given application, many bits in this expanded codeword are zero.
All immediate values are then sign-extended from the first significant digit.
For example, although the immediate offset field is 12 bits, only 4 bits may be used by the current application (this particular case arises from stack-pointer manipulation in the code of Section~\ref{sec:quicksort}).

The expanded instruction word---which can exceed 100 bits---is then collapsed to remove all redundant bits.
The resulting optimized instruction word is typically still 50 bits or more in width.
Although this is much wider than the original 32-bit RISC-V encoding, the cost of the application-specific I-MEM is usually not significantly greater than in the original case.
Assuming roughly 400 instructions for an application, the original binary is $400\times 32$ and fits in a single M20K.
If the optimized (expanded) format is 55 bits wide and requires an average of one NOP per instruction, the I-MEM grows to $800\times 55$, which still requires only three M20Ks.
In other words, NOP insertion has limited impact on I-MEM cost, partly because the programs we optimize are small to begin with.

\subsection{Effect on decode and ALU}

The decode stage, which requires approximately 100 ALMs in the baseline core, is no longer needed.
Some decoded outputs from the previous decoder---whether instructions or immediates---no longer exist in the expanded instruction word and are instead tied to zeroes.
Many ALU-stage inputs are therefore zeroed.
The advantage of the wide, shallow ALU architecture then becomes apparent: much of the ALU is optimized away during synthesis.
We show that this typically removes more than half of the logic.

The result of the process is illustrated in Figure~\ref{fig:riscv_optimize}.
The combination of I-MEM and decoder is replaced by a memory containing the expanded instructions.
An additional pipeline stage is inserted between the PC (typically a 10-bit register) and the address port of the expanded instruction memory so that pipeline behavior remains identical to the unoptimized version.

Overall processor logic size can shrink by 50\% or more.
In other words, we can construct a 32-bit RISC-V application-specific processor in the region of $\approx$200 ALMs that closes timing above 750\,MHz in a mid-range (Agilex~5) device.
The processor can retire one instruction per clock under ideal circumstances, although the average rate may be closer to 3~CPI.

In summary, the method generates an assembler-compatible, functionally identical processor that is not binary-compatible with a general-purpose RISC-V.
Run times may differ because of software hazard mitigation, but the higher clock frequency of the optimized processor will often outweigh any increase in CPI.

\section{RV32I: Quicksort}
\label{sec:quicksort}

\begin{figure*}[t]
    \centering
    \framebox{\includegraphics[scale=0.75]{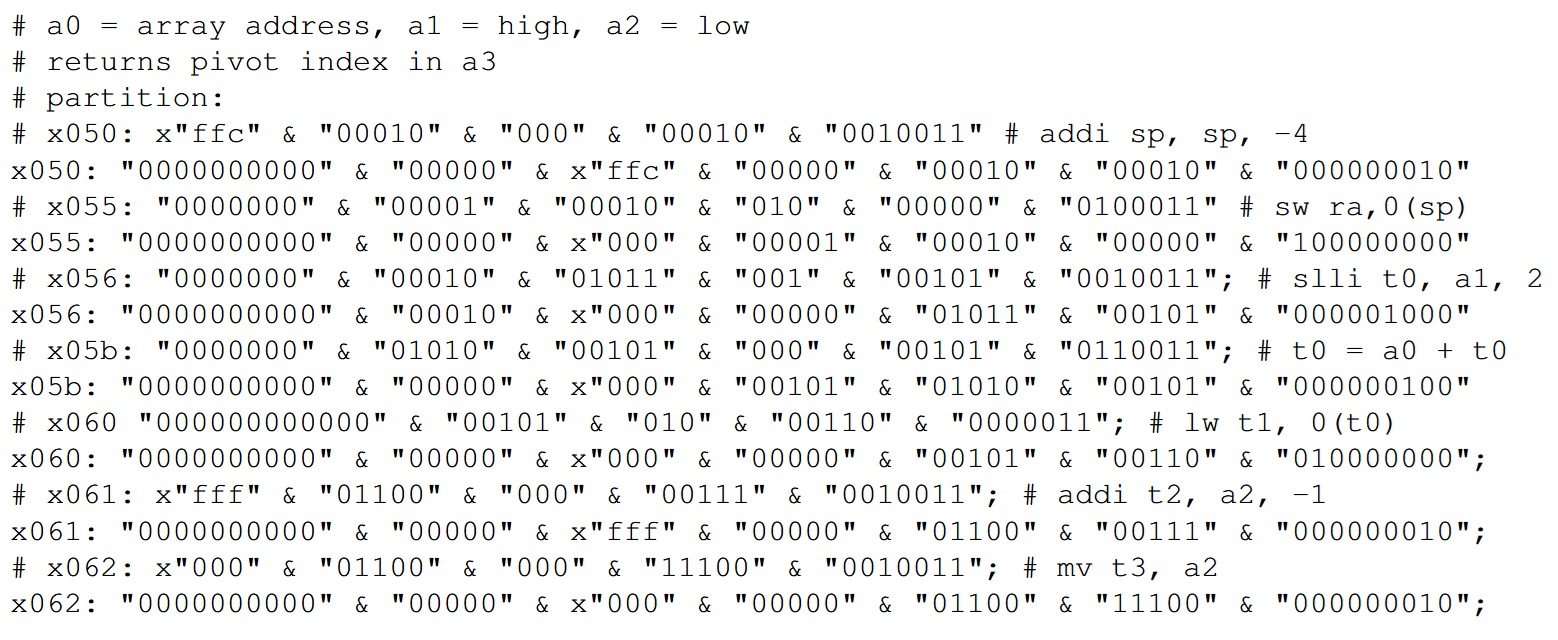}}
   \caption{Quicksort partition extract: code mapping from RV32I fields to the expanded instruction format.}
   \label{fig:quicksort_code}
\end{figure*}

We use Quicksort~\cite{Hoare62a} to demonstrate the method for an RV32I configuration.
Although the algorithm is relatively compact (20 lines of ANSI C and 55 assembly instructions in our implementation), there are two recursive calls in the main routine, yielding a fairly complex program flow.
Loop bodies are usually short, so the effects of branches are acute.
Quicksort therefore clearly demonstrates the tradeoff between processor feature reduction (no forwarding) and clock rate.

The following code is an excerpt from the start of the \texttt{PARTITION} routine:

\begin{footnotesize}
\begin{verbatim}
# partition: (offset 0x6C = 108)
0x0000006C: 0xFFC10113  # addi sp, sp, -4
0x00000070: 0x00112023  # sw ra, 0(sp)
0x00000074: 0x00259293  # slli t0, a1, 2
0x00000078: 0x00550293  # add t0, a0, t0
0x0000007C: 0x0002A303  # lw t1, 0(t0)
0x00000080: 0xFFF60393  # addi t2, a2, -1
0x00000084: 0x00060E13  # mv t3, a2
\end{verbatim}
\end{footnotesize}

Figure~\ref{fig:quicksort_code} shows the first optimization steps.
Each line contains two versions of the same instruction.
First, the RV32I instruction is shown (commented with ``\#''), with each field split according to the structure of that instruction type.
The next line is the decomposition of that instruction into the expanded instruction fields.

The order of these fields, from left to right, is:
\begin{itemize}
    \item a 10-bit branch-offset field (unused in this snippet);
    \item a 5-bit shift field;
    \item a 12-bit immediate field;
    \item three 5-bit register fields: destination, source1, and source2;
    \item a one-hot instruction field.
\end{itemize}

Sign extensions of the immediate fields have not yet been applied in the figure, so those fields appear at full width.
Only nine instruction opcodes are required for Quicksort; the one-hot encodings, from left to right, are: SW, LW, BGE, JAL, JALR, SLLI, ADD, ADDI, and LUI.
The full 10-bit branch offset is used, but only six bits of the immediate field.
Including NOPs for hazard mitigation, the total instruction counts are 151 for the 8-stage processor and 96 for the 5-stage core.

\subsection{Comparison methodology}

We compiled the optimized RISC-V core using Quartus~25.1 into an Agilex~5 A5EC013AB23AI1V device.
As baselines we use the Nios~II/f (1006 ALMs) and Nios~V/c (336 ALMs) processors from Altera, compiled into the same device.
We compare primarily against Nios~V/c---a state-machine-based, multi-cycle RISC-V RV32I core---because it is Altera's smallest vendor-supplied RISC-V soft core and therefore the most relevant baseline for our target use case: minimum-footprint embedded control on low-cost FPGAs.
A larger soft core such as Nios~V/m may reduce CPI, but at substantially higher ALM cost ($\approx$1.5K ALMs) and lower FMax (492\,MHz), moving it away from the area--delay optimum targeted here.

Applying the optimization to the lightweight RISC-V core produced a large impact on both logic and clock rate.
Logic was reduced by about 60\%.
Clock rate improved by about 20\%: although logic on the critical paths in and around the ALU is removed primarily from inside the ALU pipeline, the routing distance between the ALU and register file also shrinks as the processor contracts.

The minimum memory required for a processor instance is about six M20Ks: two for the I-MEM (the expanded instruction word is typically wider than 40 bits), two for the 2R--1W register file, and two for a $1024\times 32$ D-MEM.
A smaller footprint is sometimes possible: the register file can be implemented in a single M20K configured in true dual-port (2R--2W) mode, and if a 512-word D-MEM is sufficient then only four M20Ks are required in total.
Even so, the logic-to-memory ratio remains unbalanced relative to the fabric: in the Agilex~5E a typical M20K-to-ALM ratio is about 140.
Because an optimized processor needs so few M20Ks, very efficient placement is possible---logic can sit in a regular block immediately beside the memory column (Section~\ref{sec:results}).

\subsection{Performance results}

We ran the optimized code on both pipeline depths for three 200-element test cases: a pathological monotonically increasing sequence, a pathological monotonically decreasing sequence, and a pseudo-random sequence.
We also compared the same C code compiled for Nios~II and Nios~V/c.
Interestingly, the Nios~II C compiler produced the same number of assembly instructions (55) but a different structure; for example, the gcc RISC-V compile pushed a four-element activation frame onto the stack, while the Nios toolflow built a five-element frame.

Results are summarized in Table~\ref{tab:qs_performance}.
Our objective is not minimum execution time at any cost, but the best area--delay tradeoff for small embedded soft-processor instances on FPGA.
We therefore evaluate designs using an area--delay product (ADP),\footnote{ADP $\propto$ ALMs $\times$ Cycles $\times$ Time.} which captures the relevant Pareto frontier for this application space.
Absolute runtime remains important, but it is not the primary optimization target: larger or more feature-rich processors can always move to a different point on the area/performance curve.

The actual cost is still lower than a pure ALM-count ratio suggests, because the optimized processor can be placed in a tightly constrained regular box.
(For simplicity the ADP comparison uses ALM counts only.)
There is a small delta in instruction count between our RISC-V cores and Nios~V/c; because Nios~V/c is encrypted we could not inspect its internals to analyze this difference.
Resource utilization of the optimized RISC-V cores for Quicksort is reported in Table~\ref{tab:optimized}.

\begin{table*}[t]
  \centering
  \caption{Quicksort: Nios~II vs.\ Nios~V/c vs.\ optimized RISC-V---performance and
           normalized cost. ADP (area--delay product) is normalized to RV32I-8 Stage.}
  \label{tab:qs_performance}
  \scriptsize
  \setlength{\tabcolsep}{4pt}
  \renewcommand{\arraystretch}{1.1}
  \begin{tabular}{@{}l r r @{\hspace{12pt}} l r r r r r r r@{}}
    \toprule
    \multicolumn{3}{c}{\textbf{Core}} & \multicolumn{8}{c}{\textbf{Quicksort Results}} \\
    \cmidrule(r{8pt}){1-3} \cmidrule(l){4-11}
    \textbf{Processor} & \textbf{Speed} & \textbf{ALMs} & \textbf{Testcase} &
    \textbf{Instructions} & \textbf{Cycles} & \textbf{NOPs} & \textbf{Branch} &
    \textbf{CPI} & \textbf{Time} & \textbf{ADP} \\
    & \textbf{(MHz)} & & & & & & \textbf{Slots} & & \textbf{($\mu$s)} & \\
    \midrule
    \multirow{3}{*}{Nios II}
      & \multirow{3}{*}{445} & \multirow{3}{*}{1006}
      & Up       & 268,897 &   435,615 & \na     & \na     & 1.62 &  979 & 4.42 \\
    & & & Down     & 208,897 &   355,125 & \na     & \na     & 1.70 &  798 & 4.32 \\
    & & & P-Random &  21,471 &    36,075 & \na     & \na     & 1.68 &   81 & 3.40 \\
    \midrule
    \multirow{3}{*}{Nios V/c}
      & \multirow{3}{*}{675} & \multirow{3}{*}{336}
      & Up       & 254,309 & 1,740,161 & \na     & \na     & 6.84 & 2578 & 3.90 \\
    & & & Down     & 191,369 & 1,289,679 & \na     & \na     & 6.74 & 1910 & 3.46 \\
    & & & P-Random &  27,311 &   186,834 & \na     & \na     & 6.84 &  276 & 3.87 \\
    \midrule
    \multirow{3}{*}{RV32I-5 Stage}
      & \multirow{3}{*}{478} & \multirow{3}{*}{207}
      & Up       & 253,715 &   671,893 & 328,638 &  89,540 & 2.65 & 1406 & 1.31 \\
    & & & Down     & 171,372 &   555,581 & 256,433 & 127,776 & 3.24 & 1162 & 1.10 \\
    & & & P-Random &  25,273 &    72,000 &  29,326 &  17,401 & 2.85 &  151 & 1.30 \\
    \midrule
    \multirow{3}{*}{\textbf{RV32I-8 Stage}}
      & \multirow{3}{*}{\textbf{784}} & \multirow{3}{*}{\textbf{170}}
      & Up       & 253,715 & 1,028,275 & 617,865 & 156,695 & 4.05 & 1310 & 1.00 \\
    & & & Down     & 171,372 &   858,435 & 463,455 & 223,608 & 5.01 & 1094 & 1.00 \\
    & & & P-Random &  25,273 &   110,607 &  54,751 &  30,583 & 4.38 &  141 & 1.00 \\
    \bottomrule
  \end{tabular}
\end{table*}


\begin{table*}[t]
  \centering
  \caption{Optimized RISC-V RV32I resources and performance for Quicksort.}
  \label{tab:optimized}
  \footnotesize
  \setlength{\tabcolsep}{3pt}
  \renewcommand{\arraystretch}{1.1}
  \begin{tabular}{@{}l l @{\hspace{6pt}} c c c @{\hspace{6pt}} c c c c @{\hspace{6pt}} c@{}}
    \toprule
    & & \multicolumn{3}{c}{\textbf{ALMs}} & \multicolumn{4}{c}{\textbf{Registers}} & \\
    \cmidrule(lr{4pt}){3-5} \cmidrule(lr{4pt}){6-9}
    \textbf{Processor} & \textbf{Synthesis} &
    \textbf{Total} & \textbf{ALU} & \textbf{Delays} &
    \textbf{Total} & \textbf{Primary} & \textbf{Secondary} & \textbf{Hyper} &
    \textbf{FMax} \\
    & \textbf{Option} & & & & & & & & \textbf{(MHz)} \\
    \midrule
    \multirow{2}{*}{RISC-V 8-Stage}
      & Default            & 178 & 85 & 42 & 349 & 300 & 49 &   0 & 784 \\
      & No Reset on Delays & 170 & 87 & 21 & 349 & 248 & 16 &  85 & 775 \\
    \midrule
    \multirow{2}{*}{RISC-V 5-Stage}
      & Default            & 207 & 84 & 35 & 295 & 260 & 35 &   0 & 473 \\
      & No Reset on Delays & 207 & 84 & 12 & 292 & 182 &  9 & 101 & 478 \\
    \bottomrule
  \end{tabular}
\end{table*}

\section{RV32M: FFT}
\label{sec:fft}

We next created an RV32M variant of the lightweight processor.
We did not produce a synthesizable optimized instance of this configuration; the goal was to explore application efficiency.
The motivation becomes clear in Section~\ref{sec:rv32f}: on Agilex devices, which contain embedded floating-point operators, floating-point compute can be more efficient than integer compute.

A 32-bit integer multiplier must be constructed from two DSP blocks---one in sum-of-two $19\times 18$ mode and the other as two independent $19\times 18$ multipliers.
That arrangement requires a latency of at least four cycles to match the performance level of our lightweight processor, the same pipeline depth as a DSP block in floating-point mode.
Our FFT uses only \texttt{MULH} from the standard M extension, which reduces logic slightly while retaining a four-cycle latency.
Many floating-point applications also need an integer multiplier for address generation, but those typically use only \texttt{MUL} (the lower product bits).
Because indexes rarely exceed 64K for an application-specific processor, this can be implemented with a single $19\times 18$ multiplier: small and inexpensive.

Table~\ref{tab:fft} profiles a 256-point radix-2 FFT.
The CPI of 1.96 is much closer to that of a typical full-featured processor (with forwarding and branch prediction) because of the FFT dataflow.
There are multiple groups of memory accesses without mutual dependencies, and the butterfly dataflow can be written to reduce dependencies further.
A higher-radix butterfly would potentially expose even fewer dependencies.
Branch delay slots still account for about 14\% of cycles for the radix-2 FFT.
Many of those branches are calls to the butterfly subroutine; switching to a radix-4 FFT would cut them by about 75\%, improving both throughput and CPI.


\begin{table}[t]
  \centering
  \caption{Profile of a 256-point radix-2 FFT.}
  \label{tab:fft}
  \small
  \setlength{\tabcolsep}{3pt}
  \renewcommand{\arraystretch}{1.1}
  \begin{tabular}{@{}c c @{\hspace{6pt}} c c @{\hspace{6pt}} c c c@{}}
    \toprule
    \multicolumn{2}{c}{\textbf{Totals}} &
    \multicolumn{2}{c}{\textbf{Pipeline Overhead}} &
    \multicolumn{3}{c}{\textbf{Instruction Mix}} \\
    \cmidrule(r{4pt}){1-2} \cmidrule(lr{4pt}){3-4} \cmidrule(l{4pt}){5-7}
    \textbf{OPs} & \textbf{Cycles} &
    \textbf{NOPs} & \textbf{Branch} &
    \textbf{Load} & \textbf{Store} & \textbf{Butterfly} \\
    & & & \textbf{Slots} & & & \textbf{OPs} \\
    \midrule
    27,128 & 53,145 & 18,973 & 7224 & 5375 & 4096 & 11,774 \\
    \bottomrule
  \end{tabular}
\end{table}

\section{RV32F: Gaussian Matrix Solver}
\label{sec:rv32f}

\begin{figure}[t]
    \centering
    \includegraphics[width=\columnwidth]{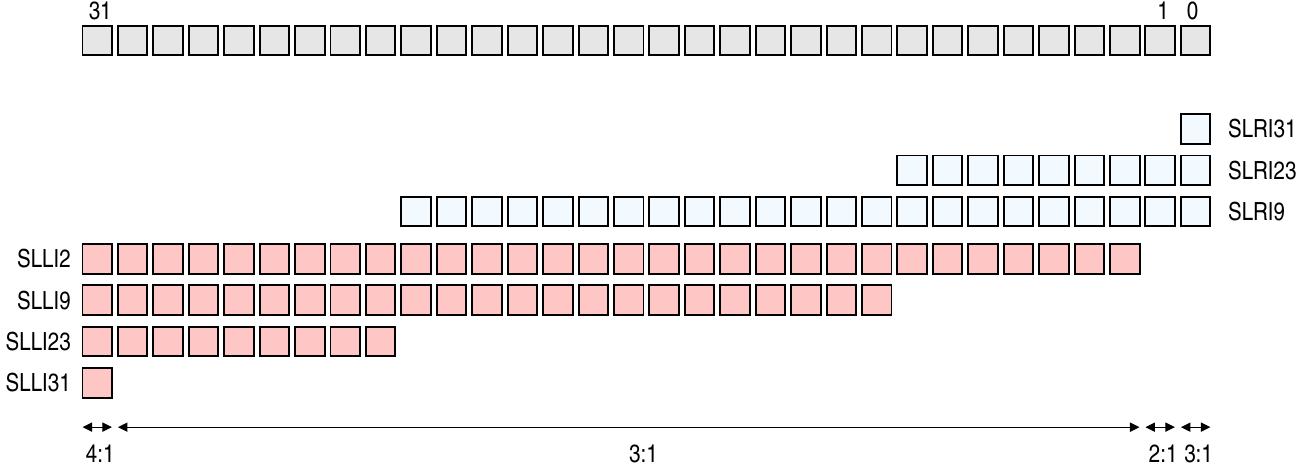}
   \caption{Enumerated shift complexity versus a full immediate shifter.}
   \label{fig:immed_shift}
\end{figure}

We then created an RV32F instance with hardware support only for FP32 multiply and add/subtract.
Other FP32 functions, such as divide and square root, are implemented in software.
This core requires somewhat more logic than the RV32I and RV32M cores: the DSP block in floating-point mode does not support dynamic configuration, so we use a DSP datapath that implements both a floating-point multiplier and an adder.
Individual functions---multiply, add, and subtract---are selected by modifying operands in soft logic before the DSP inputs.
In high-performance mode the floating-point DSP has a fixed 4-cycle latency, longer than the two-cycle latency of our 8-stage RISC-V ALU.
This required two additional NOPs to avoid hazards.
With heterogeneous ALUs, the 6-cycle round-trip hazards apply not only to FP--FP sequences but also to writeback conflicts with following integer instructions.

Our application example is a Gaussian matrix solver: the matrix is triangulated and the solution obtained by back substitution.
Elementary functions were written in C and compiled to RISC-V; the solver itself is written in assembler.
Indexing required an integer multiplier except when matrix dimensions were powers of two, in which case multipliers could be replaced by left shifts.

The profile of a $4\times 4$ matrix solution is shown in Table~\ref{tab:gauss_code}.
Of the 513 instructions, 133 were floating-point operations---i.e., the majority of instructions were used for addressing and branching.
The four floating-point divisions required 85 clock cycles each, including hazard NOP slots.
CPI was 3.54, similar to our other benchmarks.
Loop control and normalization dominate on a $4\times 4$ matrix; larger matrices would improve CPI substantially.


\begin{table}[t]
  \centering
  \caption{Profile of a $4\times 4$ Gaussian matrix solver.}
  \label{tab:gauss_code}
  \small
  \setlength{\tabcolsep}{3pt}
  \renewcommand{\arraystretch}{1.1}
  \begin{tabular}{@{}c c @{\hspace{6pt}} c c @{\hspace{6pt}} c c c@{}}
    \toprule
    \multicolumn{2}{c}{\textbf{Totals}} &
    \multicolumn{2}{c}{\textbf{Pipeline Overhead}} &
    \multicolumn{3}{c}{\textbf{Instruction Mix}} \\
    \cmidrule(r{4pt}){1-2} \cmidrule(lr{4pt}){3-4} \cmidrule(l{4pt}){5-7}
    \textbf{OPs} & \textbf{Cycles} &
    \textbf{NOPs} & \textbf{Branch} &
    \textbf{SLLI} & \textbf{SLLR} & \textbf{FP} \\
    & & & \textbf{Slots} & & & \textbf{OPs} \\
    \midrule
    513 & 1815 & 1085 & 217 & 74 & 8 & 133 \\
    \bottomrule
  \end{tabular}
\end{table}

\subsection{Enumerated immediate shifts}

Floating-point division and square root can be implemented efficiently with Newton--Raphson iteration.
These routines often require an initial estimate, which in turn requires extracting the floating-point argument's sign, exponent, and mantissa.
The following immediate shifts are used: SLLI31, SLLI23, SLLI20, SLLI9, SRLI9, SRLI23, and SRLI31.
This creates a problem for the optimization method: to the synthesis tool it appears that the entire shifter is required.
Shifters are the largest component in the processor ALU stage, and every bit position of the shift amount is used at least once for both left and right shifts.

We therefore introduce an additional optimization: all immediate shifts are enumerated as separate instructions.
This adds eight instructions in total (including the ubiquitous SLLI2 used to convert indexes into addresses).
Rather than two separate datapath structures with five levels of 2:1 multiplexers, the enumerated shift reduces to a 3:1 multiplexer for most bit positions (Figure~\ref{fig:immed_shift}).

Table~\ref{tab:gauss_resource} presents resource utilization (compiled to the A5EC013AB23AI1V device) for the processor configurations benchmarked on the Gaussian solver.
The bottom row reports the configuration with enumerated shifts.
The impact is significant: a further 25\% logic reduction over the already optimized version.
The INT27 multiplier is needed for index calculations when matrix sizes are not powers of two; the additional datapath multiplexers add about 20\% ALMs to the core.


\begin{table*}[t]
  \centering
  \caption{Optimized RISC-V F-extension resources for the Gaussian matrix solver.}
  \label{tab:gauss_resource}
  \small
  \setlength{\tabcolsep}{3pt}
  \renewcommand{\arraystretch}{1.1}
  \begin{tabular}{@{}l l @{\hspace{6pt}} c c @{\hspace{6pt}} c c c @{\hspace{6pt}} c@{}}
    \toprule
    & & \multicolumn{2}{c}{\textbf{ALMs}} & & & & \\
    \cmidrule(lr{4pt}){3-4}
    \textbf{Processor} & \textbf{Configuration} &
    \textbf{Total} & \textbf{ALU} &
    \textbf{Registers} & \textbf{M20K} & \textbf{DSP} &
    \textbf{FMax} \\
    & & & & & & & \textbf{(MHz)} \\
    \midrule
    \multirow{2}{*}{Lightweight}
      & + FP32           & 494 & 249 &  942 & 5 & 1 & 668 \\
      & + FP32 + INT27   & 591 & 250 & 1029 & 5 & 2 & 622 \\
    \midrule
    \multirow{2}{*}{Optimized}
      & Standard           & 394 & 236 & 694 & 6 & 1 & 668 \\
      & Enumerated Shifts  & 298 & 133 & 639 & 6 & 1 & 668 \\
    \bottomrule
  \end{tabular}
\end{table*}

\section{Placement Results}
\label{sec:results}

\newlength{\fpheight}
\setlength{\fpheight}{4.5cm}
\newsavebox{\fpbox}
\newcommand{\fpimg}[2]{%
  \sbox{\fpbox}{\includegraphics[height=\fpheight]{#2}}%
  \ifdim\wd\fpbox<#1\relax
    \makebox[#1][c]{\usebox{\fpbox}}%
  \else
    \usebox{\fpbox}%
  \fi}

\begin{figure*}[t]
    \centering
    \subfloat[1 core\label{fig:1_core}]{%
        \fpimg{0pt}{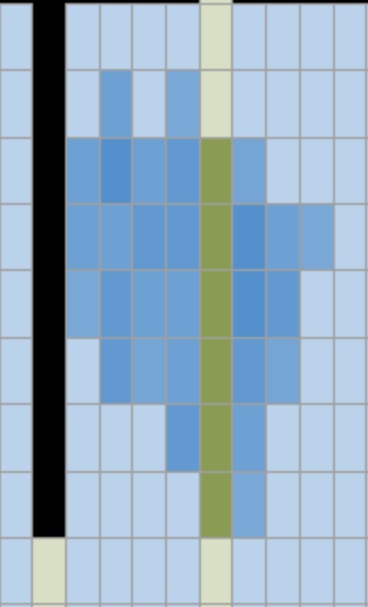}}%
    \hfill
    \subfloat[1 core constrained\label{fig:1_core_fixed}]{%
        \fpimg{3.6cm}{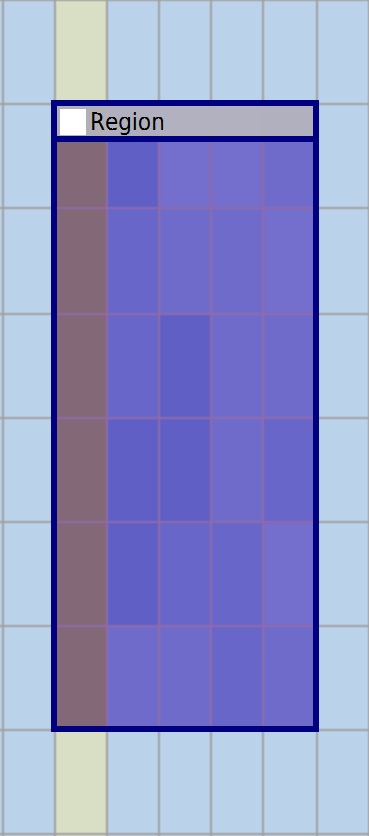}}%
    \hfill
    \subfloat[5 cores\label{fig:5_cores}]{%
        \fpimg{0pt}{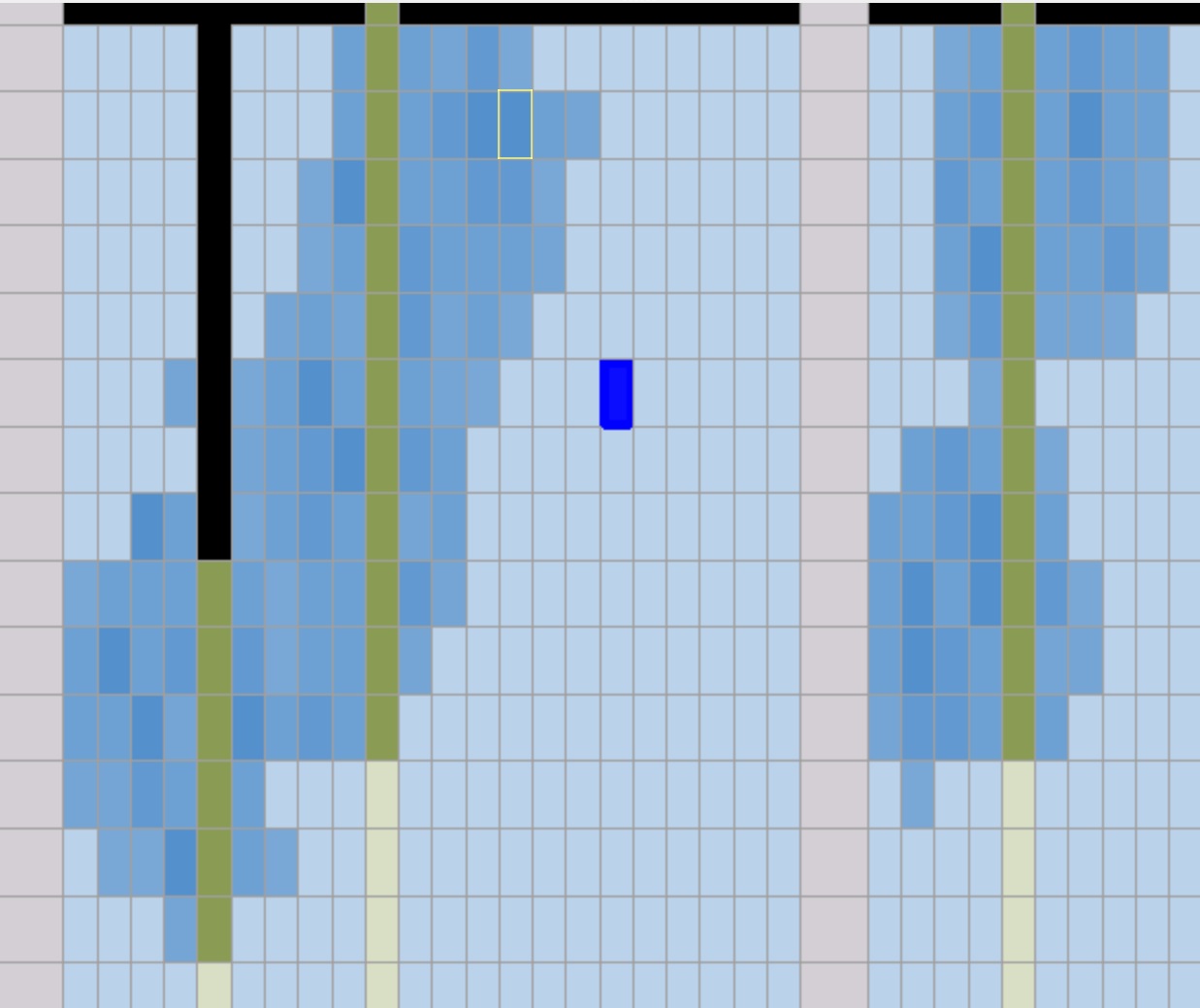}}%
    \hfill
    \subfloat[5 cores constrained\label{fig:5_cores_fixed}]{%
        \fpimg{3.6cm}{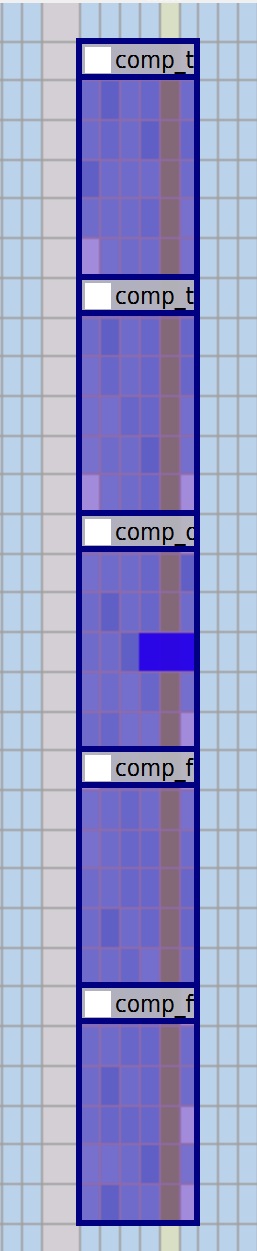}}%
    \caption{Floorplans for unconstrained and constrained implementations with 1 core and 5 cores.}
    \label{fig:floorplans}
\end{figure*}


Having reported resources, operating frequency, and algorithmic performance, we now analyze \emph{footprint}---the actual cost of the processors, including multi-processor implementations---on the FPGA.

Figure~\ref{fig:1_core} shows an unconstrained compile into an A5EC013AB23AI1V device.
Logic is centered around a column of six M20K memories (I-MEM, D-MEM, and register files).
If placement is tightly constrained, as in Figure~\ref{fig:1_core_fixed}, the entire processor logic fits in a $6\times 4$ LAB box (a Logic Array Block is a group of 10 ALMs).
In this example the logic is forced entirely to one side of the M20K column; the optimized processor does not need to straddle the memory column to achieve high performance.

An unconstrained compile of five cores is shown in Figure~\ref{fig:5_cores}.
A step-and-repeat vertical stacking of the single-core constraint produces the structure in Figure~\ref{fig:5_cores_fixed}.
Both unconstrained compiles closed timing at 784\,MHz; both constrained examples closed at 724\,MHz.
The latter result gives confidence that systems with deterministic area and performance can be realized in practice.

\section{Conclusions}
\label{sec:conclusions}

Soft processors have a long history on FPGAs, but a full processor system is often unnecessary---especially on smaller devices.
Compiling the final processor instance only after application development is complete can offer a significant cost reduction.
Cost is more than a resource-list count: it is the footprint of the design on the FPGA.

We have shown that a standard CPU---here RISC-V---can be reduced to a fraction of a full-featured implementation by specializing the instruction encoding and datapath to a known application binary.
Beyond resource reduction, the small area improves timing closure and enables packing many processors together in a step-and-repeat manner.
The maximum clock potential of the FPGA can often be achieved for these optimized cores, even in multi-processor constructions, making it possible to plan system density and performance deterministically.

\section*{Acknowledgments}
The authors thank colleagues at Altera Corporation for discussions on soft-processor placement and high-frequency design methodology.

\bibliographystyle{IEEEtran}
\bibliography{references.bib}

\end{document}